\documentclass[proceedings, preprint]{rmaa}

\SetYear{2010}
\SetConfTitle{XIII Latin American Regional IAU Meeting}

\title{An Overabundance of Oxygen in Planetary Nebulae of the Solar
Neighborhood}

\author{
  M. Rodr\'\i guez,\altaffilmark{1} 
  and G. Delgado-Inglada\altaffilmark{1}}

\altaffiltext{1}{Instituto Nacional de Astrof\'\i sica,
  \'Optica y Electr\'onica (INAOE), Apdo Postal 51 y 216, 72000 Puebla,
  Mexico (mrodri@inaoep.mx, gloria@inaoep.mx)}

\shortauthor{Rodr\'\i guez \& Delgado-Inglada}
\shorttitle{Oxygen in the Solar Neighborhood}

\listofauthors{M. Rodr\'\i guez \& G. Delgado-Inglada}
\indexauthor{Rodr\'\i guez, M.}
\indexauthor{Delgado-Inglada, G.}

\abstract{We study the oxygen abundance in five \ion{H}{2} regions and seven
planetary nebulae (PNe) located within 2 kpc from the Sun that have available
spectra of high quality. Our analysis uses a similar procedure and the same
atomic data to derive abundances in all the objects. The results calculated with
collisionally excited lines for the \ion{H}{2} regions indicate that the local
interstellar medium is very homogeneous, with
$12+\log(\mbox{O}/\mbox{H})=8.45\mbox{--}8.54$. As for the PNe, six out of seven
show significantly higher abundances:
$12+\log(\mbox{O}/\mbox{H})=8.65\mbox{--}8.80$. This overabundance of oxygen in
PNe also holds when we consider the abundances implied by recombination lines.}

\resumen{Estudiamos la abundancia de ox\'\i geno en cinco regiones \ion{H}{2} y
siete nebulosas planetarias (NPs) situadas a menos de 2 kpc de distancia del sol
que tienen espectros disponibles de alta calidad. Nuestro an\'alisis usa
un procedimiento similar y los mismos datos at\'omicos para calcular las
abundancias en todos los objetos. Los resultados encontrados con l\'\i neas de
excitaci\'on colisional para las regiones
\ion{H}{2} indican que el medio interestelar local es muy homog\'eneo, con
$12+\log(\mbox{O}/\mbox{H})=8.45\mbox{--}8.54$. En cuanto a las NPs, seis de
las siete muestran abundancias significativamente mayores:
$12+\log(\mbox{O}/\mbox{H})=8.65\mbox{--}8.80$. Esta sobreabundancia de ox\'\i
geno en las NPs se mantiene cuando consideramos las abundancias que implican
las l\'\i neas de recombinaci\'on.}

\addkeyword{H~II regions}
\addkeyword{ISM: abundances}
\addkeyword{planetary nebulae: general}

\begin{document}
\maketitle

\section{Introduction}

We present a comparison between the oxygen abundances derived homogeneously for
ionized gas in \ion{H}{2} regions and planetary nebulae (PNe) of the solar
neighborhood. This approach is the only way in which we can compare the
abundances of the interstellar medium (ISM, represented in this context by
\ion{H}{2} regions) with stellar abundances (represented by PNe) using the same
procedure and the same atomic data. Furthermore, the abundance analysis is based
on optically thin lines, providing a large advantage over abundances derived
directly for stars. We choose oxygen for this
analysis because it is the element for which we can get the most reliable
abundances in ionized gas. Besides, the oxygen abundance is not expected to be
substantially modified by the evolutionary processes that take place in the
progenitors of PNe with near-solar metallicity (although the most massive
progenitors can achieve a small amount of destruction -- see, e.g.,
\citealp{kar10}).

In principle, our approach has one disadvantage. Weak recombination lines (RLs)
of heavy elements imply higher abundances than collisionally excited lines
(CELs), by factors around or above 2, both in \ion{H}{2} regions and PNe (e.g.,
\citealp{gar06,liu00}). To avoid this difficulty, we will consider here the
results implied by both RLs and CELs. However, different explanations of the
discrepancy imply that the best abundance estimates will be close to the ones
implied by either RLs or CELs, or will be intermediate between them (see
\citealp{rod10}, and references therein). Hence, the interpretation of the
results must take into account that the explanation of the discrepancy can be
different for \ion{H}{2} regions and PNe.

Our sample contains five \ion{H}{2} regions (M8, M16, M17, M20, and M42) and
seven PNe (NGC~3132, NGC~3242, NGC~6210, NGC~6543, NGC~6572, NGC~6720, and
NGC~6884) with deep spectra (\citealp{est04,gar06,gar07,liu04,tsa03,wes04}). All
the objects have individual distance determinations locating them at distances
below 2 kpc. This constraint should minimize the effects of the Galactic
abundance gradient. The PNe were selected from the sample of low-ionization
nebulae compiled by \citet{del09}, and are estimated to have
$\mbox{O}^{3+}/\mbox{O}<0.15$. Since no [\ion{O}{4}] lines are observed in the
optical along with all the other lines we will be using, this restriction
reduces the uncertainties introduced by the correction for the presence of
O$^{3+}$.

\section{Results}

All the objects were analyzed in a similar way and using the same atomic data.
An average electron density, $n_e$, was obtained using 2--3 diagnostic line
ratios. Electron temperatures, $T_e$, for the low- and high-ionization regions
inside each nebula were obtained from the usual [\ion{N}{2}] and [\ion{O}{3}]
diagnostic line ratios (see, e.g., \citealp{ost06}). The $\mbox{O}^{+}/\mbox{H}^+$ and
$\mbox{O}^{++}/\mbox{H}^+$ abundance ratios were calculated using the values of
$n_e$ and $T_e$ ($T_e$[\ion{N}{2}] for O$^+$; $T_e$[\ion{O}{3}] for O$^{++}$)
and the line intensity ratios $I([\ion{O}{2}]~\lambda3727)/I(\mbox{H}\beta)$ and
$I([\ion{O}{3}]~\lambda\lambda4959,5007)/I(\mbox{H}\beta)$. The total oxygen
abundance was derived by adding the O$^{+}$ and O$^{++}$ abundances, and using
ionization correction factors for the presence of O$^{3+}$ that go from
$\mbox{O}/(\mbox{O}^{+}+\mbox{O}^{++})=1$ (for
the \ion{H}{2} regions and some PNe) to 1.18. For the results implied by RLs,
the O$^{++}$ abundance was derived using the \ion{O}{2} RLs of multiplet 1.
Then, we estimated the total oxygen abundance implied by RLs by assuming the
same ionization fractions found with CELs.
Further details on the procedure will be provided in Rodr\'\i guez \&
Delgado-Inglada (2011, in preparation).

\begin{figure}[!t]
  \includegraphics[bb=60 270 553 683,clip,width=7.9cm]{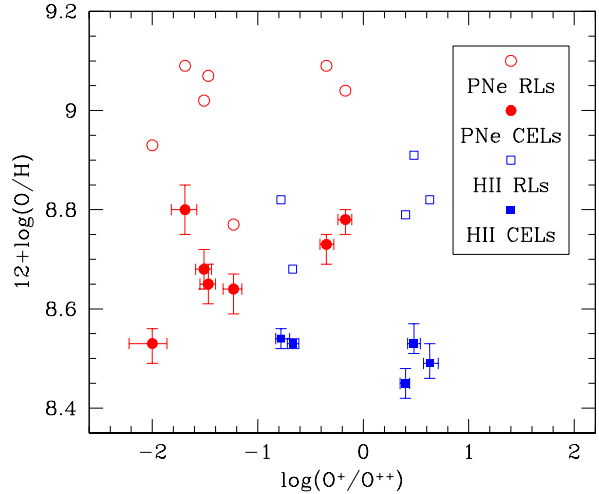}
  \caption{Oxygen abundances implied by CELs (filled symbols) and RLs (open
  symbols) as a function of $\mbox{O}^+/\mbox{O}^{++}$ for \ion{H}{2} regions
  (squares) and PNe (circles) in the solar neighborhood. The error bars
  show the observational uncertainties.}
  \label{fig:uno}
\end{figure}

Figure~\ref{fig:uno} shows the oxygen abundances implied by CELs and RLs as a
function of $\mbox{O}^+/\mbox{O}^{++}$ for all the objects in the sample. The
abundances implied by CELs in \ion{H}{2} regions are similar, with
$12+\log(\mbox{O}/\mbox{H})=8.45\mbox{--}8.54$, suggesting that the local ISM is
very homogeneous. As for the PNe, if we exclude NGC~3242 (where the correction
for O$^{3+}$ is the largest, making its total oxygen abundance more uncertain),
the abundances implied by CELs are in the range
$12+\log(\mbox{O}/\mbox{H})=8.65\mbox{--}8.80$. Since the oxygen abundances of
PNe are expected to reflect the abundances in the ISM from which their
progenitor stars formed several gigayears ago, this result is opposite to what
we would predict using simple chemical evolution models. This overabundance of
oxygen in PNe also holds when we consider the abundances derived with RLs. In
fact, since the abundances implied by RLs in \ion{H}{2} regions are similar to
the abundances implied by CELs in PNe, even results intermediate between those
implied by CELs and RLs (and shifted by different amounts in both kinds of
objects) would indicate the presence of an overabundance in PNe.

This overabundance of oxygen in PNe could be due to different causes, like
oxygen production in the stellar progenitors, large-scale gas flows in the
Galaxy, recent infall of low-metallicity gas,
or extensive stellar migration from the inner Galaxy. 
A discussion of these possibilities and a comparison of the results with those
implied by stars of different ages and by those based on absorption lines in the
diffuse ISM will be presented elsewhere (Rodr\'\i guez \& Delgado-Inglada 2011,
in preparation).

\end{document}